\documentclass[english]{article}
\usepackage[T1]{fontenc}
\usepackage[utf8]{inputenc}
\usepackage{array}
\usepackage{multirow}
\usepackage{amsmath}
\usepackage{amsthm}
\usepackage{amssymb}
\usepackage{rotfloat}
\usepackage{subscript}

\usepackage{nicefrac}
\usepackage{tikz}

\usepackage{xcolor}

\makeatletter

\providecommand{\tabularnewline}{\\}

\theoremstyle{plain}
\newtheorem{thm}{\protect\theoremname}

\@ifundefined{date}{}{\date{}}
\usepackage{adjustbox}
\usepackage{amssymb}

\makeatother

\usepackage{babel}
\providecommand{\theoremname}{Theorem}

\begin{document}
\title{A note on the relation between the Contextual Fraction and CNT\textsubscript{2}}
\author{Víctor H. Cervantes\thanks{University of Illinois at Urbana-Champaign}}
\maketitle
\begin{abstract}
Contextuality (or lack thereof) is a property of systems of random
variables. Among the measures of the degree of contextuality, two
have played important roles. One of them, Contextual Fraction ($\text{CNTF}$)
was proposed within the framework of the sheaf-theoretic approach
to contextuality, and extended to arbitrary systems in the Contextuality-by-Default
approach. The other, denoted $\text{CNT}_{2}$, was proposed as one
of the measures within the Contextuality-by-Default approach. In this
note, I prove that $\text{CNTF}=2\text{CNT}_{2}$ within a class of
systems, called cyclic, that have played a prominent role in contextuality
research.
\end{abstract}
Contextuality (or lack thereof) is a property of systems of random
variables. The theory of contextuality, which in psychology has developed
from the extension of the theory of selective influences, provides
the means to detect this property and, more recently, to measure the
degree of (non)contextu\-ality of a system. Among several measures of
the degree of contextuality that have been proposed in the literature,
I consider here two: the Contextual Fraction (\(\text{CNTF}\)), proposed
by Abramsky, Barbosa, and Mansfeld
\cite{Abramsky.2017.Contextual},
and \(\text{CNT}_{2}\), proposed by Kujala and Dzhafarov \cite{Kujala.2019.Measures}.
Both have played prominent roles.
\(\text{CNTF}\) has been associated
with the quantification of the computational advantages of some quantum
computation processes \cite{Abramsky.2017.Contextual}.
On the other hand, \(\text{CNT}_{2}\) has been shown to be the only measure of contextuality
proposed so far that can be extended to a measure of noncontextuality,
\(\text{NCNT}_{2}\) \cite{Kujala.2019.Measures,Dzhafarov.2020.Contextuality}.
Moreover, \(\text{NCNT}_{2}\) is the only measure of noncontextuality
currently available.

In psychology, several applications of contextuality analysis
are due to the emergence of the quantum cognition field
(see e.g.,
\cite{Bruza.2009.Introduction,Busemeyer.2012.Quantum,Busemeyer.2015.What,Khrennikov.2014.Quantumc,Bruza.2015.Quantum, Dzhafarov.2017.Quantum}),
whereas other applications come from the selective influences literature.
In quantum cognition,
the search for contextuality has mainly served as means to
advocate for the appropriateness of the quantum formalism.
For example, in the study of language, the presence of contextuality
may witness the lack compositionality of concepts
\cite{Bruza.2015.probabilistic}.
As an extension of the theory of selective influences,
the aim has been to develop a principled approach to the ubiquitous violations
of marginal selectivity in psychological research (e.g.,
\cite{Basieva.2019.True,Cervantes.2017.Exploration,Cervantes.2018.Snow,Cervantes.2019.True,Dzhafarov.2012.Quantuma,Dzhafarov.2015.there,Dzhafarov.2016.contextuality,Dzhafarov.2017.Probability,Zhang.2015.Noncontextuality}).
Almost all of these applications include experimental data or designs
whose representation is in a prominent class of systems of random
variables, called cyclic systems (defined below). Dzhafarov, Kujala,
and Cervantes \cite{Dzhafarov.2020.Contextuality,Dzhafarov.2021.Epistemic}
have completely characterized the measure \(\text{CNT}_{2}\) for cyclic
systems. Using their results, I present a proof that, in the class
of cyclic systems, \(\text{CNTF} = 2\text{CNT}_{2}\).

\section{Contextuality-by-Default}

This section introduces the definitions and results from the Contextuality-by-Default
theory (CbD)
\cite{Dzhafarov.2016.Contextuality-by-Default:,Dzhafarov.2017.Contextuality-by-Default,Dzhafarov.2016.Context-content,Kujala.2019.Measures,Dzhafarov.2019.Contextuality-by-Default,Dzhafarov.2020.Contextuality}
that are needed for the proof.
Throughout the text, I shall use the
data from Cervantes and Dzhafarov \cite{Cervantes.2018.Snow} to illustrate
the defined objects and computations.
Their data come from an online experiment in which participants were asked to make
two conjoint choices, one of a character and one of a characteristic to describe the character,
in accordance with a storyline sketching Hans Christian Andersen's
``The Snow Queen''.
The manipulations determined the pair of characters,
and the pair of characteristics from which a participant made their
choices.

A \emph{system} of random variables is a set of double-indexed random
variables \(R_{q}^{c}\), where \(q \in Q\) denotes their \emph{content},
which can be interpreted as a question to which the random variable
responds, and \(c \in C\) is their \emph{context}, the conditions under
which it is recorded. A system can be presented as
\begin{equation}
\mathcal{R} = \left\{ R_{q}^{c} : c \in C, q \in Q, q \prec c \right\}, \label{eq:generalSystem}
\end{equation}
where \(q \prec c\) denotes that content \(q\) is recorded in context \(c\).
The sets of contents and contexts for the ``Snow Queen'' experiment,
\(Q_{\text{SQ}} = \left\{ q_{1}, q_{2}, q_{3}, q_{4} \right\} \) and
\(C_{\text{SQ}} = \left\{ c_{1}, c_{2}, c_{3}, c_{4} \right\} \),
are presented in Tables \ref{fig:SnowQueenContexts} and \ref{fig:SnowQueenContents},
and \(\mathcal{R}_{\text{SQ}}\) will denote the corresponding system
of random variables.

For each \(q\in Q\), the subset of the system
\begin{equation}
\mathcal{R}_{q} = \left\{ R_{q}^{c} : c\in C, q\prec c \right\} \label{eq:connection}
\end{equation}
is referred to as the \emph{connection} corresponding to content \(q\).
No two random variables within a connection \(\mathcal{R}_{q}\) are
jointly distributed;
thus, they are said to be \emph{stochastically unrelated}.\footnote{More generally,
any two \(R_{q}^{c}, R_{q'}^{c'} \in \mathcal{R}\) with
\(c \neq c'\) are stochastically unrelated.}
Analogously, for each \(c \in C\), the subset
\begin{equation}
R^{c} = \left\{ R_{q}^{c} : q \in Q, q\prec c \right\} \label{eq:bunch}
\end{equation}
is referred to as the \emph{bunch} for context \(c\).
The variables within a bunch are jointly distributed.
That is, bunches are random vectors with a given probability distribution.
Table \ref{tab:bunchesSQ} presents the joint distributions of
the four bunches of \(\mathcal{R}_{\text{SQ}}\).

\begin{table}
\caption{Contents of the ``Snow Queen'' experiment.}
\label{fig:SnowQueenContexts}
\centering{}%
\begin{tabular}{ccc}
\hline
Content & \multicolumn{2}{c}{Choice between} \tabularnewline
\hline
\(q_{1}\) & Gerda          & Troll \tabularnewline
\(q_{2}\) & Beautiful      & Unattractive \tabularnewline
\(q_{3}\) & The Snow Queen & Old Finn woman \tabularnewline
\(q_{4}\) & Kind           & Evil \tabularnewline
\hline
\end{tabular}
\end{table}

\begin{table}
\caption{Contexts of the ``Snow Queen'' experiment.}
\label{fig:SnowQueenContents}
\centering{}%
\begin{tabular}{cc}
\hline
Context & Conjoint choices made \tabularnewline
\hline
\(c_{1}\) & \(q_{1}\) and \(q_{2}\) \tabularnewline
\(c_{2}\) & \(q_{2}\) and \(q_{3}\) \tabularnewline
\(c_{3}\) & \(q_{3}\) and \(q_{4}\) \tabularnewline
\(c_{4}\) & \(q_{4}\) and \(q_{1}\) \tabularnewline
\hline
\end{tabular}
\end{table}

\begin{table}
\caption{Joint distributions of each of the bunches of the ``Snow Queen'' experiment.}
\label{tab:bunchesSQ}

\begin{adjustbox}{max width = \textwidth}
\begin{tabular}{lrrrclrrr}
\hline
& \multicolumn{2}{c}{\(R_{2}^{1}\)} & Mar. &&
& \multicolumn{2}{c}{\(R_{4}^{4}\)} & Mar. \tabularnewline
\cline{2-3} \cline{7-8}
\multicolumn{1}{c}{\(R^{c_{1}}\)} & Beautiful (0) & Unattractive (1) & character &&
\multicolumn{1}{c}{\(R^{c_{4}}\)} & Kind (0)      & Evil (1)         & character \tabularnewline
\hline
\(R_{1}^{1}\)     &          &          &          &&
\(R_{1}^{4}\)     &          &          &          \tabularnewline
\(\ \ \)Gerda (0) & \(.843\) & \(.020\) & \(.864\) &&
\(\ \ \)Gerda (0) & \(.797\) & \(.035\) & \(.832\) \tabularnewline
\(\ \ \)Troll (1) & \(.029\) & \(.107\) & \(.136\) &&
\(\ \ \)Troll (1) & \(.018\) & \(.150\) & \(.168\) \tabularnewline
Mar.\ characteristic & \(.872\) & \(.128\) &  &&
Mar.\ characteristic & \(.815\) & \(.185\) & \tabularnewline
\hline
& \multicolumn{2}{c}{\(R_{2}^{2}\)} & Mar. &&
& \multicolumn{2}{c}{\(R_{4}^{3}\)} & Mar. \tabularnewline
\cline{2-3} \cline{7-8}
\multicolumn{1}{c}{\(R^{c_{2}}\)} & Beautiful (0) & Unattractive (1) & character &&
\multicolumn{1}{c}{\(R^{c_{3}}\)} & Kind (0)      & Evil (1)         & character \tabularnewline
\hline
\(R_{3}^{2}\)          &          &          &          &&
\(R_{3}^{3}\)          &          &          &          \tabularnewline
\(\ \ \)Snow Queen (0) & \(.769\) & \(.011\) & \(.780\) &&
\(\ \ \)Snow Queen (0) & \(.135\) & \(.536\) & \(.671\) \tabularnewline
\(\ \ \)Finn woman (1) & \(.070\) & \(.150\) & \(.220\) &&
\(\ \ \)Finn Woman (1) & \(.320\) & \(.009\) & \(.329\) \tabularnewline
Mar.\ characteristic   & \(.839\) & \(.161\) &  &&
Mar.\ characteristic   & \(.455\) & \(.545\) & \tabularnewline
\hline
\multicolumn{9}{l}{{Note 1. Mar. = marginal observed proportion}}\tabularnewline
\multicolumn{9}{l}{{Note 2. Adapted from Table 3 of \cite{Cervantes.2018.Snow}.
The value encoding each choice for the variables \(R_{j}^{i}\) is}} \tabularnewline
\multicolumn{9}{l}{{shown in parentheses.}} \tabularnewline
\end{tabular}
\end{adjustbox}
\end{table}

The class of systems known as cyclic has had a prominent role in contextuality research.
Cyclic systems are the object of Bell's theorem \cite{Bell.1964.Einstein,Bell.1966.problem},
the Leggett--Garg theorem \cite{Leggett.1985.Quantum},
and Suppes and Zanotti's theorem \cite{Suppes.1981.When},
among many other theoretical results (e.g., \cite{Araujo.2013.All,Kujala.2016.Proof}).
It is also encountered in most applications which empirically probe for contextuality
(e.g., \cite{Adenier.2017.Test,Asano.2014.Violation,Cervantes.2018.Snow,Dzhafarov.2015.there,Hensen.2015.Loophole-free,Wang.2013.Quantum}).
Moreover, it has been shown that a system without cyclic subsystems
is necessarily noncontextual \cite{Dzhafarov.2020.Contextuality,Abramsky2013}.
A system \(\mathcal{R}\) is \emph{cyclic} if
\begin{enumerate}
\item each of its contexts contains two jointly distributed \emph{binary} random variables,
\item each content is measured in two contexts, and
\item there is no proper subsystem of \(\mathcal{R}\) that satisfies 1 and 2.
\end{enumerate}
A cyclic system with \(n\) contexts (and \(n\) contents) is said to
be a cyclic system of \emph{rank} \(n\).
For any cyclic system, the contexts and contents can be rearranged and numbered so that the system
can be written as
\begin{equation}
\mathcal{R}_{n} = \left\{ \left\{ R_{i}^{i}, R_{i \oplus 1}^{i} \right\} : i = 1, \ldots, n \right\} ,\label{eq:cyclicSystem}
\end{equation}
where \(R_{j}^{i}\) stands for \(R_{q_{j}}^{c_{i}}\), and \(\oplus 1\)
denotes cyclic shift \(1 \mapsto 2, \ldots, n-1 \mapsto n, n \mapsto 1\).\footnote{Similarly,
\(\ominus 1\) will denote the inverse shift of \(\oplus 1\).}
Thus, the variables \(\left\{ R_{i}^{i}, R_{i \oplus 1}^{i} \right\}\)
constitute the bunch corresponding to context \(c_{i}\). The following
matrices depict three cyclic systems: a cyclic system of rank 2 (where
\emph{rank} refers to the number of contents, or, equivalently, the
number of contexts), a cyclic system of rank 3, and a cyclic system
of rank 4. The latter gives the structure of the system \(\mathcal{R}_{SQ}\)
that represents the ``Snow Queen'' experiment.
\begin{equation}
\begin{array}{ccccc}
\begin{array}{|c|c||c|}
\hline R_{1}^{1} & R_{2}^{1} & c_{1} \\
\hline R_{1}^{2} & R_{2}^{2} & c_{2} \\
\hline
\hline q_{1}     & q_{2}     & \mathcal{R}_{2} \\
\hline
\end{array}
&  &
\begin{array}{|c|c|c||c|}
\hline R_{1}^{1} & R_{2}^{1} &           & c_{1} \\
\hline           & R_{2}^{2} & R_{3}^{2} & c_{2} \\
\hline R_{1}^{3} &           & R_{3}^{3} & c_{3} \\
\hline
\hline q_{1}     & q_{2}     & q_{3}     & \mathcal{R}_{3} \\
\hline
\end{array}
&  &
\begin{array}{|c|c|c|c||c|}
\hline R_{1}^{1} & R_{2}^{1} &           &           & c_{1} \\
\hline           & R_{2}^{2} & R_{3}^{2} &           & c_{2} \\
\hline           &           & R_{3}^{3} & R_{4}^{3} & c_{3} \\
\hline R_{1}^{4} &           &           & R_{4}^{4} & c_{4} \\
\hline
\hline q_{1}     & q_{2}     & q_{3}     & q_{4}     & \mathcal{R}_{4} \\
\hline
\end{array}
\end{array}
\end{equation}

Without loss of generality, we assume that the binary random variables
in our systems take the values \(0\) and \(1\).
Table \ref{tab:bunchesSQ} shows which choice of character and of characteristic is encoded as
\(0\) and which as \(1\) for the description and contextuality computations
for the ``Snow Queen'' experiment.
Kujala and Dzhafarov \cite{Kujala.2019.Measures}
introduce some notation to produce a vectorial description of such a system.
This representation is obtained by taking the probabilities
of events for each random variable, for each bunch, and for each connection
in the system. Let \(\mathbf{l}_{(.)}\), \(\mathbf{b}_{(.)}\), and \(\mathbf{c}_{(.)}\)
denote these vectors, respectively. For describing a cyclic system,
the first two vectors are written:
\begin{equation}
\mathbf{l}_{(.)} =  \left(
\begin{array}{c}
\Pr(R_{i}^{i} = r_{i}^{i}) \\
\Pr(R_{i \oplus 1}^{i} = r_{i \oplus 1}^{i})
\end{array}
\right)_{i = 1, \ldots, n}, \label{eq:fullMargins}
\end{equation}
\begin{equation}
\mathbf{b}_{(.)} =  \left(
\begin{array}{c}
\Pr(R_{i}^{i} = r_{i}^{i}, R_{i \oplus 1}^{i} = r_{i \oplus 1}^{i})
\end{array}
\right)_{i = 1, \ldots, n}, \label{eq:fullBunches}
\end{equation}
with \(r_{i}^{i}, r_{i \oplus 1}^{i} = 0,1.\)

The vector \(\mathbf{c}_{(.)}\) is composed of imposed probabilities.
These probabilities describe a \emph{coupling} of the variables within each connection.
A coupling of a set of random variables \(\left\{ X_{i} \right\}_{i \in I}\),
where \(I\) indexes the variables in the set, is a new set of jointly
distributed random variables \(\left\{ Y_{i} \right\}_{ i\in I}\) such that for each \(i\in I\),
the distributions of \(X_{i}\) and \(Y_{i}\) coincide.
In CbD, the couplings chosen for each connection are their multimaximal couplings.
In a \emph{multimaximal coupling} of a set of random variables,
for any two \(Y_{i}, Y_{i'}\), the probability \(\Pr(Y_{i} = Y_{i'})\) is
the maximal possible given their individual distributions.
If we denote the variables of the multimaximal coupling of connection \(\mathcal{R}_{q_{j}}\) by
\begin{equation}
\mathcal{T}_{q_{j}} = \left\{ T_{j}^{i} : c_{i} \in C, q_{j} \prec c_{i} \right\}, \label{eq:connectionCoupling}
\end{equation}
one obtains the vector
\begin{align}
\mathbf{c}_{(.)} & =\left(
\begin{array}{c}
\Pr(T_{i}^{i} = r_{i}^{i}, T_{i}^{i \ominus 1} = r_{i}^{i \ominus 1})
\end{array}
\right)_{i = 1, \ldots, n}, \label{eq:fullConnection}
\end{align}
with \(r_{i}^{i}, r_{i \oplus 1}^{i} = 0, 1\), and where
\[
\begin{array}{ll}
\Pr(T_{i}^{i} = 0, T_{i}^{i \ominus 1} = 0) & = \min(\Pr(R_{i}^{i} = 0), \Pr(R_{i}^{i \ominus 1} = 0)), \\
\Pr(T_{i}^{i} = 1, T_{i}^{i \ominus 1} = 1) & = \min(\Pr(R_{i}^{i} = 1), \Pr(R_{i}^{i \ominus 1} = 1)), \\
\Pr(T_{i}^{i} = 0, T_{i}^{i \ominus 1} = 1) & = \Pr(R_{i}^{i \ominus 1} =1 ) - \Pr(T_{i}^{i} = 1, T_{i}^{i \ominus 1} = 1), \\
\text{and} \\
\Pr(T_{i}^{i}= 1 , T_{i}^{i \ominus 1} = 0) & =\Pr(R_{i}^{i} = 1) - \Pr(T_{i}^{i} = 1, T_{i}^{i \ominus 1} = 1).
\end{array}
\]
Using this notation, Table \ref{tab:connectionsSQ} presents the multimaximal
couplings for each of the four connections of system \(\mathcal{R_{\text{SQ}}}\).

\begin{table}
\caption{Distributions of the multimaximal couplings of each of the connections
of the ``Snow Queen'' experiment.}
\label{tab:connectionsSQ}
\begin{adjustbox}{max width = \textwidth}
\begin{tabular}{lcccclccc}
\hline
& \multicolumn{2}{c}{\(T_{1}^{4}\)} & Mar. &&
& \multicolumn{2}{c}{\(T_{2}^{2}\)} & Mar. \tabularnewline
\cline{2-3}  \cline{7-8}
\multicolumn{1}{c}{\(T_{q_{1}}\)} & Gerda (0)     & Troll (1)        & character &&
\multicolumn{1}{c}{\(T_{q_{2}}\)} & Beautiful (0) & Unattractive (1) & character \tabularnewline
\hline
\(T_{1}^{1}\) &  &  &  &&
\(T_{2}^{1}\) &  &  & \tabularnewline
\(\ \ \)Gerda (0)        & \(.832\) & \(.032\) & \(.864\) &&
\(\ \ \)Beautiful (0)    & \(.839\) & \(.033\) & \(.872\) \tabularnewline
\(\ \ \)Troll (1)        & \(.000\) & \(.136\) & \(.136\) &&
\(\ \ \)Unattractive (1) & \(.000\) & \(.128\) & \(.128\) \tabularnewline
Mar.\ characteristic     & \(.832\) & \(.168\) &  &&
Mar.\ characteristic     & \(.839\) & \(.161\) & \tabularnewline
\hline
& \multicolumn{2}{c}{\(T_{4}^{4}\)} & Mar. &&
& \multicolumn{2}{c}{\(T_{3}^{2}\)} & Mar. \tabularnewline
\cline{2-3}  \cline{7-8}
\multicolumn{1}{c}{\(T_{q_{4}}\)} & Kind (0) & Evil (1) & character &&
\multicolumn{1}{c}{\(T_{q_{3}}\)} & Snow Queen (0) & Finn Woman (1) & character \tabularnewline
\hline
\(T_{4}^{3}\) &  &  &  &&
\(T_{3}^{3}\) &  &  & \tabularnewline
\(\ \ \)Kind (0)       & \(.455\) & \(.000\) & \(.455\) &&
\(\ \ \)Snow Queen (0) & \(.671\) & \(.000\) & \(.671\) \tabularnewline
\(\ \ \)Evil (1)       & \(.360\) & \(.185\) & \(.545\) &&
\(\ \ \)Finn Woman (1) & \(.109\) & \(.220\) & \(.329\) \tabularnewline
Mar.\ characteristic   & \(.815\) & \(.185\) &  &&
Mar.\ characteristic   & \(.780\) & \(.220\) & \tabularnewline
\hline
\multicolumn{9}{l}{Note 1. Mar. = marginal observed proportion} \tabularnewline
\multicolumn{9}{l}{Note 2. The variable \(T_{j}^{i}\) denotes
the variable in the coupling corresponding to \(R_{j}^{i}\).} \tabularnewline
\end{tabular}
\end{adjustbox}
\end{table}

The system \(\mathcal{R}_{SQ}\) is then described by vectors\footnote{The third probability
in vector \(\mathbf{b}_{(.)}^{*}\) has been modified to \(.021\) to adjust for rounding error,
so that the four shown values of the first context precisely sum to \(1\).}
\[
\begin{split}
\mathbf{l}_{(.),SQ}^{*\intercal} = & (.864, .136, .872, .128, .839, .161, .780, .220, \\
&  .671, .329, .455, .545, .815, .185, .832, .168),
\end{split}
\]
\[
\begin{split}
\mathbf{b}_{(.),SQ}^{*\intercal} = & (.843, .029, .021, .107, .769, .011, .070, .150, \\
&  .135, .320, .536, .009, .797, .035, .018, .150),
\end{split}
\]
\[
\begin{split}
\mathbf{c}_{(.),SQ}^{*\intercal} = & (.832, .000, .032, .136, .839, .033, .000, .128, \\
&  .671, .109, .000, .220, .455, .000, .360, .185).
\end{split}
\]

A system is noncontextual if it is possible to construct a coupling
of \(\mathcal{R}\) that agrees simultaneously with the bunches and
with the multimaximal couplings of the connections
\cite{Dzhafarov.2016.Contextuality-by-Default:,Dzhafarov.2017.Contextuality-by-Default}.
If such a coupling exists, it can be found as the solution to a system
of linear equations defined using vectors \(\mathbf{l}_{(.)}\), \(\mathbf{b}_{(.)}\),
and \(\mathbf{c}_{(.)}\).
First, let \(\mathbf{p}_{(.)}\) be the concatenation
of the three vectors \(\mathbf{l}_{(.)}\), \(\mathbf{b}_{(.)}\), and \(\mathbf{c}_{(.)}\).
Next, consider a vector \(\mathbf{s}\) of length \(2^{2n}\) of all possible values
that a coupling of the entire system \(\mathcal{R}_{n}\) can take.
That is, if we denote a coupling of \(\mathcal{R}_{n}\) by \(S_{n}\),
then \(S_{n}\) is a random vector whose values are the conjunction of events
\[
\left\{ S_{j}^{i} = r_{j}^{i} : i, j = 1, \ldots, n, \ q_{j} \prec c_{i} \right\},
\]
with \(r_{j}^{i} = 0, 1\).
These values are the components of \(\mathbf{s}\).
Let \(\mathbf{M}_{(.)}\) be an incidence \((0/1)\) matrix with \(2^{2n}\)
columns labeled by the elements of \(\mathbf{s}\) and \(12n\) rows labeled
by the events whose probabilities are the components of \(\mathbf{p}_{(.)}\).
This matrix is constructed as follows:
\begin{itemize}
\item If the \(u\)th component of \(\mathbf{p}_{(.)}\) is \(\Pr(R_{j}^{i} = r_{j}^{i}\))
and the \(v\)th component of \(\mathbf{s}\) includes the event
\(\left\{ S_{j}^{i} = r_{j}^{i} \right\}\),
then the cell \((u, v)\) of \(\mathbf{M}_{(.)}\) is a \(1\);
\item if the \(u\)th component of \(\mathbf{p}_{(.)}\) is
\(\Pr(R_{i}^{i} = r_{i}^{i}, R_{i \oplus i}^{i} = r_{i \oplus 1}^{i})\)
and the \(v\)th component of \(\mathbf{s}\) includes the event
\(\left\{ S_{i}^{i} = r_{i}^{i}, S_{i \oplus 1}^{i} = r_{i \oplus 1}^{i} \right\}\),
then the cell \((u,v)\) is a \(1\);
\item if the \(u\)th component of \(\mathbf{p}_{(.)}\) is
\(\Pr(T_{i}^{i} = r_{i}^{i}, T_{i}^{i \ominus 1} = r_{i}^{i \ominus 1})\)
and the \(v\)th component of \(\mathbf{s}\) includes the event
\(\left\{ S_{i}^{i} = r_{i}^{i}, S_{i}^{i \ominus 1} = r_{i}^{i \ominus 1} \right\}\),
then the cell \((u,v)\) is a \(1\);
\item all other cells have zeroes.
\end{itemize}
A detailed description of the construction of \(\mathbf{M}_{(.)}\)
for general systems of binary random variables can be found in Dzhafarov and Kujala
\cite{Dzhafarov.2016.Context-content}.

The system \(\mathcal{R}_{n}\) described by \(\mathbf{p_{(.)}^{*}}\),
is noncontextual \cite{Dzhafarov.2016.Context-content} if and only
if there is a vector \(\mathbf{h}\geq0\) (component-wise) such that
\begin{equation}
\mathbf{M}_{(.)} \mathbf{h} = \mathbf{p_{(.)}^{*}}. \label{eq:largeLFT}
\end{equation}
Any solution \(\mathbf{h}^{*}\) provides the probability distribution
of a coupling \(S_{n}\) of \(\mathcal{R}_{n}\) that contains as its
marginals both the bunch distributions and the multimaximal couplings
of the connections of \(\mathcal{R}_{n}\).

It is evident that the rows of \(\mathbf{M}_{(.)}\) are not linearly
independent. For example, one can compute
\[
\Pr(R_{i}^{i} = 0) + \Pr(R_{i}^{i} = 1) = 1,
\]
and
\[
\begin{array}{cc}
\Pr(R_{i}^{i} = 0, R_{i \oplus 1}^{i} = 0) + \Pr(R_{i}^{i} = 0, R_{i \oplus 1}^{i} = 1) \\
+ \Pr(R_{i}^{i} = 1, R_{i \oplus 1}^{i} = 0) + \Pr(R_{i}^{i} = 1, R_{i \oplus 1}^{i} = 1) = & 1.
\end{array}
\]
Hence, the row of \(\mathbf{M}_{(.)}\) corresponding to \(\Pr(R_{i}^{i} = 0)\)
is a linear combination of the rows corresponding to
\(\Pr(R_{i}^{i} = 0, R_{i \oplus 1}^{i} = 0)\),
\(\Pr(R_{i}^{i} = 0, R_{i \oplus 1}^{i} = 1)\),
\(\Pr(R_{i}^{i} = 1, R_{i \oplus 1}^{i} = 0)\),
\(\Pr(R_{i}^{i} = 1, R_{i \oplus 1}^{i} = 1)\),
and
\(\Pr(R_{i}^{i} = 1)\).

The vectorial description can be reduced by taking only a subset of
the components of \(\mathbf{p}_{(.)}\),
such that the corresponding rows of \(\mathbf{M}_{(.)}\) are linearly independent.
Let us choose the following reductions of the vectors
\(\mathbf{l}_{(.)}\), \(\mathbf{b}_{(.)}\), and \(\mathbf{c}_{(.)}\)
and denote them
\(\mathbf{l}\), \(\mathbf{b}\), and \(\mathbf{c}\):
\begin{equation}
\label{eq:reducedVectors}
\begin{array}{cllc}
\mathbf{l} & = \left( p_{j}^{i} \right)_{i, j = 1, \ldots, n, \ q_{j} \prec c_{i}}
& = \left( \Pr(R_{j}^{i} = 1) \right)_{i, j = 1, \ldots, n, \ q_{j} \prec c_{i}}, \\
\mathbf{b} & = \left( p_{i, i \oplus 1} \right)_{i = 1, \ldots, n}
& = \left( \Pr(R_{i}^{i} = 1, R_{i \oplus 1}^{i} = 1) \right)_{i = 1 , \ldots, n},
&   \text{and} \\
\mathbf{c} & = \left( p^{i, i \ominus 1} \right)_{i = 1, \ldots, n}
& = \left( \Pr(T_{i}^{i} = 1, T_{i}^{i \ominus 1} = 1) \right)_{i = 1, \ldots, n}.
\end{array}
\end{equation}
The system \(\mathcal{R}_{\text{SQ}}\) can then be described by the
reduced vectors
\[
\begin{array}{clc}
\mathbf{l}_{SQ}^{*\intercal} & = (.136, .128, .161, .220, .329, .545, .185, .168), \\
\mathbf{b}_{SQ}^{*\intercal} & = (.107, .150, .009, .150), & \text{and} \\
\mathbf{c}_{SQ}^{*\intercal} & = (.136, .128, .220, .185).
\end{array}
\]

Using the reduced vectors (\ref{eq:reducedVectors}), Kujala and Dzhafarov
\cite{Kujala.2019.Measures} define the linear programming tasks used
to compute the CbD-based measures of contextuality.
Let a system \(\mathcal{R}_{n}\) be described by
\begin{equation}
\mathbf{p}^{*} = \left(
\begin{array}{c}
\mathbf{l^{*}} \\
\mathbf{b^{*}} \\
\mathbf{c^{*}}
\end{array}
\right),
\end{equation}
where \(\mathbf{l^{*}}\) and \(\mathbf{b^{*}}\) are the empirical probabilities of the system,
and \(\mathbf{c}^{*}\) are the probabilities found from the multimaximal couplings of each of its connections.
Let \(\mathbf{M}\) be the incidence matrix found by taking the rows of \(\mathbf{M}_{(.)}\)
corresponding to the elements of \(\mathbf{p}^{*}\).
Note that the system is noncontextual if and only if there is a vector \(\mathbf{h}\geq0\)
(component-wise) such that
\begin{equation}
\mathbf{Mh} = \mathbf{p}^{*}, \label{eq:LFT}
\end{equation}
subject to \(\mathbf{1}^{\intercal} \mathbf{h} = 1\) \cite{Dzhafarov.2016.Context-content}.
Denoting the rows of \(\mathbf{M}\) that correspond to \(\mathbf{l}^{*}\), \(\mathbf{c}^{*}\), \(\mathbf{b}^{*}\)
by, respectively, \(\mathbf{M_{l}}\), \(\mathbf{M_{b}}\), \(\mathbf{M_{c}}\),
we can rewrite (\ref{eq:LFT}) in extenso:
\begin{equation}
\label{eq:extensoLFT}
\left(
\begin{array}{c}
\mathbf{M_{l}} \\
\mathbf{M_{b}} \\
\mathbf{M_{c}}
\end{array}
\right)
\mathbf{h} = \left(
\begin{array}{c}
\mathbf{l^{*}} \\
\mathbf{b^{*}} \\
\mathbf{c^{*}}
\end{array}
\right).
\end{equation}
For the system \(\mathcal{R}_{\text{SQ}}\) that describes the data
from the ``Snow Queen'' experiment, the matrix \(\mathbf{M}\) and
the corresponding submatrices \(\mathbf{M_{l}}\), \(\mathbf{M_{b}}\),
and \(\mathbf{M_{c}}\) are depicted in Table \ref{fig:matrixM} in
the Appendix.

The contextuality measure \(\text{CNT}_{2}(\mathcal{R}_{n})\) can then
be computed solving the linear programming task
\begin{equation}
\label{lp:CNT2}
\begin{tabular}{|ccc|}
\hline
find           & minimizing                            & subject to \tabularnewline
\hline
\(\mathbf{x}\) & \(\mathbf{1}^{\intercal} \mathbf{d}\) &
\(-\mathbf{d} \leq \mathbf{b}^{\ast} - \mathbf{M}_{\mathbf{b}} \mathbf{x} \leq \mathbf{d}\) \tabularnewline
& & \(\mathbf{x}, \mathbf{d} \geq 0\)         \tabularnewline
& & \(\mathbf{1}^{\intercal} \mathbf{x} = 1\) \tabularnewline
& & \(\mathbf{M_{l} x} = \mathbf{l}^{\ast}\)  \tabularnewline
& & \(\mathbf{M_{c} x} =\mathbf{c}^{\ast}\)   \tabularnewline
\hline
\end{tabular}.
\end{equation}
For any solution \(\mathbf{x^{*}}\),
\(\text{CNT}_{2}(\mathcal{R}_{n}) = \left\Vert \mathbf{b}^{*} - \mathbf{M}_{\mathbf{b}} \mathbf{x^{*}} \right\Vert _{1}
= \left\Vert \mathbf{p}^{*} - \mathbf{M} \mathbf{x^{*}} \right\Vert _{1}\).
\cite{Dzhafarov.2020.Contextuality} showed that
for a cyclic system, the value of \(\text{CNT}_{2}(\mathcal{R})\) is
a one-coordinate distance; that is, if \(\mathcal{R}_{n}\) is contextual,
there exists a solution \(\mathbf{x}^{*}\) such that \(\mathbf{d}^{*}\)
has zeroes for all but one of its components.
They further showed that for each coordinate \(i = 1,\ldots,n\) of \(\mathbf{d}\),
there is a solution \(\mathbf{x}^{*}\) such that \(d_{i}\) is the nonzero component.
Note that there are infinitely many solutions to the linear programming
task, since there are at least two solutions. Table \ref{fig:solutionsSQ}
in the Appendix presents one such solution to this task for system
\(\mathcal{R}_{\text{SQ}}\) representing the ``Snow Queen'' experiment.
The contextuality measure \(\text{CNT}_{2}\) of the system is \(.069\).

A system is said to be \emph{consistently connected} if, for any two
\(R_{q}^{c}, R_{q}^{c'}\), the respective distributions of \(R_{q}^{c}\)
and \(R_{q}^{c'}\) coincide.
This property of a system corresponds to the \emph{marginal selectivity}
condition in the selective influences literature in psychology,
as well as to the \emph{no-disturbance}
or the \emph{no-signaling} requirements in quantum physics.
If this property is not satisfied, the system is said to be \emph{inconsistently connected.}
In general, a system of random variables \(\mathcal{R}\)
is not consistently connected. Note that whenever a system \(\mathcal{R}\)
is consistently connected, then for any two \(R_{q}^{c}, R_{q}^{c'}\),
the corresponding variables of a multimaximal coupling of \(\mathcal{R}_{q}\)
are almost always equal (that is, \(\Pr(T_{q}^{c} = T_{q}^{c'}) = 1\)).

The contextuality measure \(\text{CNTF}\) proposed in \cite{Abramsky.2017.Contextual}
is defined only for consistently connected systems. However, there
is a procedure that can be applied to any system of binary random
variables to generate a new system \(\mathcal{R}^{\ddagger}\) that
is consistently connected and whose contextually status is the same
as that of the original system \(\mathcal{R}\).
Moreover, if \(\mathcal{R}\) is consistently connected, then \(\text{CNTF}(\mathcal{R})\)
and \(\text{CNTF}(\mathcal{R}^{\ddagger})\) are equal
\cite{Dzhafarov.2019.Contextuality-by-Default}.
The \emph{consistification} of system \(\mathcal{R}\) is obtained by constructing a new system
\(\mathcal{R}^{\ddagger}\) in the following manner.
First, define the set of contents \(Q^{\ddagger}\) of the new system as
\[
Q^{\ddagger} = \left\{ q_{ij} : c_{i} \in C, q_{j} \in Q, q_{j} \prec c_{i} \right\} .
\]
That is, for each content \(q_{j}\) and each of the contexts \(c_{i}\)
in which it is recorded, we define a content \(q_{ij}\)=``\(q_{j}\)
recorded in context \(c_{i}\)''.
Next, define the new set of contexts \(C^{\ddagger}\) as
\[
C^{\ddagger} = C \sqcup Q,
\]
the disjoint union of the contexts and the contents of the system \(\mathcal{R}\).
Then, define the new relation
\[
\prec^{\ddagger} = \left\{ (q_{ij}, c_{i}) : q_{j} \in Q, c_{i} \in C, q_{j} \prec c_{i} \right\}
\sqcup
\left\{ (q_{ij}, q_{j}) : q_{j} \in Q, c_{i} \in C, q_{j} \prec c_{i}\right\}.
\]
That is, the new content \(q_{ij}\) is recorded in precisely two of
the new contexts, \(c_{i} \in C^{\ddagger}\) and \(q_{j} \in C^{\ddagger}\).
In this manner, the bunch
\[
R^{c_{i}}=\left\{ R_{q_{ij}}^{c_{i}}:q_{ij}\in Q^{\ddagger},q_{ij}\prec^{\ddagger}c_{i}\right\}
\]
coincides with the bunch
\[
R^{c_{i}} = \left\{ R_{q}^{c_{i}} : q \in Q, q \prec c_{i} \right\}
\]
of the original system; while the bunch
\[
R^{q_{j}} = \left\{ R_{q_{ij}}^{q_{j}} : q_{ij} \in Q^{\ddagger}, q_{ij} \prec^{\ddagger} q_{j} \right\}
\]
is constructed by defining new jointly distributed random variables
\(\left\{ R_{q_{ij}}^{q_{j}} \right\}_{q_{ij} \prec^{\ddagger} q_{j}}\)
such that \(R^{q_{j}}\) is the multimaximal coupling of \(\mathcal{R}^{q_{j}}\).
Note that if \(\mathcal{R}\) is a cyclic system of rank \(n\), then
its consistified system \(\mathcal{R}^{\ddagger}\) is a consistently
connected cyclic system of rank \(2n\).
The following matrices show the consistification of system \(\mathcal{R}_{2}\)
and how its bunches relate to the bunches of the original system and the multimaximal
couplings of its connections.
\begin{equation}
\begin{array}{ccc}
\begin{array}{|c|c|c|c||c|}
\hline
R_{q_{11}}^{c_{1}} & R_{q_{12}}^{c_{1}} &                    &                    & c_{1} \\
\hline
& R_{q_{12}}^{q_{2}} & R_{q_{22}}^{q_{2}} &                    & q_{2} \\
\hline
&                    & R_{q_{22}}^{c_{2}} & R_{q_{21}}^{c_{2}} & c_{2} \\
\hline
R_{q_{11}}^{q_{1}} &                    &                    & R_{q_{21}}^{q_{1}} & q_{1} \\
\hline\hline
q_{11}             & q_{12}             & q_{22}             & q_{21}             & \mathcal{R}_{2}^{\ddagger} \\
\hline
\end{array}
&  &
\begin{array}{|c|c|c|c||c|}
\hline
R_{1}^{1} & R_{2}^{1} &           &           & c_{1} \\
\hline\hline
& T_{2}^{1} & T_{2}^{2} &           & q_{2} \\
\hline\hline
&           & R_{2}^{2} & R_{1}^{2} & c_{2} \\
\hline\hline
T_{1}^{1} &           &           & T_{1}^{2} & q_{1} \\
\hline
\end{array}
\end{array}
\end{equation}

If a system \(\mathcal{R}\) is consistently connected, the contextual
fraction proposed by Abramsky et al. \cite{Abramsky.2017.Contextual}
can be computed solving the following linear programming task:
\begin{equation}
\label{lp:CNTF}
\begin{tabular}{|ccc|}
\hline
find           & maximizing                            & subject to \tabularnewline
\hline
\(\mathbf{z}\) & \(\mathbf{1}^{\intercal} \mathbf{z}\) &
\(\mathbf{M}_{(.)} \mathbf{z} \leq \mathbf{p}_{(.)}^{\ast}\) \tabularnewline
& & \(\mathbf{z} \geq 0\)                                        \tabularnewline
& & \(\mathbf{1}^{\intercal} \mathbf{z} \leq 1\)                 \tabularnewline
\hline
\end{tabular}.
\end{equation}
For any solution \(\mathbf{z^{*}}\), \(\text{CNTF}(\mathcal{R}) = 1 - \mathbf{1^{\intercal}z^{*}}\).
As with the task to compute \(\text{CNT}_{2}\), there are infinitely
many solutions to this task.
The previous task is equivalent to the
one proposed in \cite{Abramsky.2017.Contextual} which uses a simpler
representation of the system \cite{Dzhafarov.2019.Contextuality-by-Default}.
Now, if we consider the consistified system \(\mathcal{R}_{n}^{\ddagger}\)
of a cyclic system \(\mathcal{R}_{n}\), then
equality \ref{eq:cnt2equality} is satisfied by Th. 18 of \cite{Dzhafarov.2020.Contextuality} and,
if \(\mathcal{R}_{n}\) is consistently connected,
equality \ref{eq:cntfequality} is satisfied by Th. 7 of \cite{Dzhafarov.2019.Contextuality-by-Default}.
\begin{align}
\text{CNT}_{2}(\mathcal{R}_{n}) & =\text{CNT}_{2}(\mathcal{R}_{n}^{\ddagger}), \label{eq:cnt2equality} \\
\text{CNTF}(\mathcal{R}_{n})    & =\text{CNTF}(\mathcal{R}_{n}^{\ddagger}). \label{eq:cntfequality}
\end{align}
Moreover, Th. 7 of \cite{Dzhafarov.2019.Contextuality-by-Default}
also shows that, regardless of consistent connectedness, the linear
programming task to compute \(\text{CNTF}(\mathcal{R}^{\ddagger})\)
is equivalent to the task in expression (\ref{lp:CNTF}) where \(\mathbf{p}^{*}_{(.)}\)
describes system \(\mathcal{R}^{\ddagger}\).
Hence, we will abuse notation and simply write \(\text{CNT}_{2}\) and
\(\text{CNTF}\) for the measures of the systems \(\mathcal{R}_{n}\)
and \(\mathcal{R}_{n}^{\ddagger}\), and use equality (\ref{eq:cntfequality})
as the definition of the contextual fraction for inconsistently connected
systems.
Table \ref{fig:solutionsSQ}
presents one solution to the task (Expression~\ref{lp:CNTF}) for the system
\(\mathcal{R}_{\text{SQ}}\).
The contextuality measure \(\text{CNTF}\) of the system is \(.138\).

Lastly, we will need the definition of the
\emph{noncontextuality polytope} and the related
objects defined by
\cite{Dzhafarov.2020.Contextuality} to describe its geometry.
The noncontextuality polytope is usually defined
with respect to the expected values of the variables codified
with outcomes \(-1/1\) and their products, rather than
with respect to the probabilities.
These expected values can be obtained by transforming the components of vectors
\(\mathbf{l}\), \(\mathbf{b}\), and \(\mathbf{c}\) as follows
\begin{align}
    \label{eq:expectedLow}
         e_{j}^{i}         &=   2 p_{j}^{i} - 1, & q_{j} \prec c_{i}, \\
    \label{eq:expectedBunch}
         e_{i, i \oplus 1} &=   4 p_{i, i \oplus 1} - 2 p_{i}^{i} - 2 p_{i \oplus 1}^{i} + 1, &i = 1, \ldots, n, \\
    \label{eq:expectedConnection}
         e^{i, i \ominus 1} &=   4 p^{i, i \ominus 1} - 2 p_{i}^{i} - 2 p_{i}^{i \ominus 1} + 1, &i = 1, \ldots, n.
\end{align}
The resulting  vectors will be denoted as%
\begin{equation}
\label{eq:expectationVectors}
\begin{array}{cllc}
\phi(\mathbf{l}) & = \left( e_{j}^{i} \right)_{i, j = 1, \ldots, n, \ q_{j} \prec c_{i}}, \\
\phi(\mathbf{b}) & = \left( e_{i, i \oplus 1} \right)_{i = 1, \ldots, n}, \\
&   \text{and} \\
\phi(\mathbf{c}) & = \left( e^{i, i \ominus 1} \right)_{i = 1, \ldots, n}.
\end{array}
\end{equation}
Given vectors \(\phi(\mathbf{l}^{*})\) and \(\phi(\mathbf{c}^{*})\),
the noncontextuality polytope is the set
\begin{equation}
    \label{eq:noncontextuality-polytope}
    \mathbb{E}_{\mathbf{b}} =
        \left\{
            \phi(\mathbf{b}) :
            \exists \mathbf{h}
            \text{ s.t. }
            \left(
                \begin{array}{c}
                    \mathbf{M_{l}} \\
                    \mathbf{M_{b}} \\
                    \mathbf{M_{c}}
                \end{array}
            \right)
            \mathbf{h} = \left(
            \begin{array}{c}
                \mathbf{l^{*}} \\
                \mathbf{b}\phantom{^{*}} \\
                \mathbf{c^{*}}
            \end{array}
            \right)
        \right\}.
\end{equation}
That is, the noncontextuality polytope contains the set of
expected values of the products of the two variables in each bunch
that are consistent with the marginal distributions
and the implied multimaximal couplings of a given system.

The polytope \(\mathbb{E}_{\mathbf{b}}\) is a subset of the
\(n\)-cube
\begin{equation}
    \label{eq:nCube}
    \mathbb{C}_{\mathbf{b}} = [-1, 1]^{n},
\end{equation}
which consists of all the possible values of \(\phi(\mathbf{b})\).
Given an arbitrary \(n\)-box
\begin{equation}
    \label{eq:nBox}
    \mathbb{B} =
    \prod_{i = 1}^{n} [\min x_{i}, \max x_{i}] \subseteq \mathbb{C}_{\mathbf{b}},
\end{equation}
a vertex \(V\) of \(\mathbb{B}\) will be called \emph{odd} if its
coordinates contain an odd number of \(\min x_{i}\) and \emph{even}, otherwise.
In particular, a vertex of \(\mathbb{C}_{\mathbf{b}}\) is even when
the product of its coordinates equals \(1\).
\cite{Dzhafarov.2020.Contextuality} showed that
the polytope \(\mathbb{E}_{\mathbf{b}}\) is the intersection of two subsets
of \(\mathbb{C}_{\mathbf{b}}\).
For a consistently connected cyclic system, these sets are
\(\mathbb{D}_{\mathbf{b}}\), defined as
the \(n\)-demicube on the \emph{even} vertices of \(\mathbb{C}_{\mathbf{b}}\)%
---that is, the convex hull of the even vertices%
\footnote{Every \(n\)-cube can generate two \(n\)-demicubes.
One the \(n\)-demicube described in-text.
The other is the covex hull of its odd vertices.}
of  the \(n\)-cube \(\mathbb{C}_{\mathbf{b}}\)---%
and the \(n\)-box
\begin{equation}
    \label{eq:nAmbient}
    \mathbb{R}_{\mathbf{b}} =
    \prod_{i = 1}^{n}
        \left[    |e_{i}^{i} + e_{i \oplus 1}^{i}| - 1,
              1 - |e_{i}^{i} - e_{i \oplus 1}^{i}|
        \right].
\end{equation}

\section{Main theorem\label{sec:proof}}

\begin{figure}[tp]
\definecolor{naranja}{rgb}{0.90, 0.30, 0.15}
\definecolor{azul}{rgb}{0.08, 0.16, 0.30}

\def\epsilon{.05}

    \caption{Illustration of the proof for a consistently connected system \(\mathcal{R}_{2}\).}

    \begin{center}
    \begin{tikzpicture}[scale = 3.5]
    %
            \draw [line width = .7pt, gray]
                (-1, -1) -- (-1, 1) -- (1, 1) -- (1, -1) -- cycle;
            \node [] at (1 - 1.5*\epsilon, -1 + 1.5*\epsilon) {\(\mathbb{C}_{\mathbf{b}}\)};
    %
            \draw [line width = .7pt, gray, fill = naranja, fill opacity = .25]
                (-.8, -.8) -- (-.8, .8) -- (.8, .8) -- (.8, -.8) -- cycle;
            \node [azul] at (.8 - 1.5*\epsilon, -.8 + 1.5*\epsilon) {\(\mathbb{R}_{\mathbf{b}}\)};
    %
             \draw [line width = .7pt, azul]
                 (-1, -1) -- (1, 1);
    %
             \draw [line width = .7pt, azul, dashed]
                 (-1, -.6) -- (.6, 1);
    %
            \draw [line width = .7pt, azul, opacity = .7, stealth-stealth]
                (-1,  1 - .45*\epsilon) -- (.5 - .25 *\epsilon, .5 - .25*\epsilon)
                node [below = 3pt, midway] {\(\bar{C}\)};
             \draw [line width = .7pt, azul, dotted]
                 (-1,  1) -- (.2, .6);
             \draw [line width = .7pt, azul, dotted, stealth-stealth]
                 (.2, .6) -- (.5, .5)
                node [above = 2pt, near end] {\(\bar{B}\)};
    %
        \draw [line width = .7pt, azul, opacity = .7, stealth-stealth]
            (-1.01, -1) -- (-1.01, 1)
            node [left = 1pt, near end] {\(\bar{L}\)};

            \draw [line width = .7pt, azul, dotted, stealth-stealth]
                 (.2, .6) -- (.2, .2)
                node [left = 2pt, midway] {\(\bar{A}\)};
            \draw [line width = .7pt, azul, dotted, stealth-stealth]
                 (-1, -1) -- (-1, -.6)
                node [right = 2pt, near end] {\(\bar{A}'\)};
    %
            \node [] at (-1 - \epsilon, 1 + \epsilon) {\(V^{*}\)};
            \draw [line width = .7pt, gray, stealth-] (-1, -1) -- (-1,  1);
            \draw [line width = .7pt, gray, -stealth] (-1,  1) -- ( 1,  1);
            \node [fill = gray, draw, circle, inner sep = 0pt] at (-1,  1) {.};
    %
            \node [] at (.2 + \epsilon, .2 - \epsilon) {\(\mathbf{x}^{'}\)};
            \node [fill = gray, draw, circle, inner sep = 0pt] at (.2, .2) {.};
    %
            \node [] at (-1 - \epsilon, -.6) {\(\mathbf{y}\)};
            \node [fill = gray, draw, circle, inner sep = 0pt] at (-1, -.6) {.};
    %
            \node [] at (.5 + 2.0*\epsilon, .5 - 1.0*\epsilon) {\(\mathbf{x}^{NC}\)};
            \node [fill = gray, draw, circle, inner sep = 0pt] at (.5, .5) {.};
    %
            \node [fill = gray, draw, circle, inner sep = 0pt] at (.2, .6) {.};
            \node [pin={[pin distance = .5cm]90:\(\phi(\mathbf{b}^{*})\)}] at (.2, .6) {.};
    %
            \node [] at (-1 - 1.5*\epsilon, -1 - \epsilon) {\tiny{\((-1, -1)\)}};
            \node [] at ( 1 + 1.5*\epsilon, -1 - \epsilon) {\tiny{\(( 1, -1)\)}};
            \node [] at ( 1 + 1.5*\epsilon,  1 + \epsilon) {\tiny{\(( 1,  1)\)}};
    %
            \node [below = +3pt] at (0, -1 - 1.5*\epsilon) {\(e_{1,2}\)};
            \node [left]  at (-1 - 2.5*\epsilon, 0) {\(e_{2,1}\)};
    \end{tikzpicture}
    \end{center}
    \label{fig:polytope}
    \footnotesize{\raggedright{Note: Example of the noncontextuality polytope and the contextuality
    measures for a consistently connected
    cyclic
    system
    of rank 2
    with marginal expectations
    \(e_{1}^{1} = e_{1}^{2} = -.2\) and \(e_{2}^{1} = e_{2}^{2} = 0\).
    The 2-demicube \(\mathbb{D}_{\mathbf{b}}\) on the even vertices of
    \(\mathbb{C}_{\mathbf{b}}\) is the diagonal line between the points
    \((-1, -1)\) and \((1, 1)\).
    The noncontextuality polytope \(\mathbb{E}_{\mathbf{b}}\) is the segment
    of that diagonal that intersects the box \(\mathbb{R}_{\mathbf{b}}\).
    See the text for the remaining elements of the figure.}}
\end{figure}
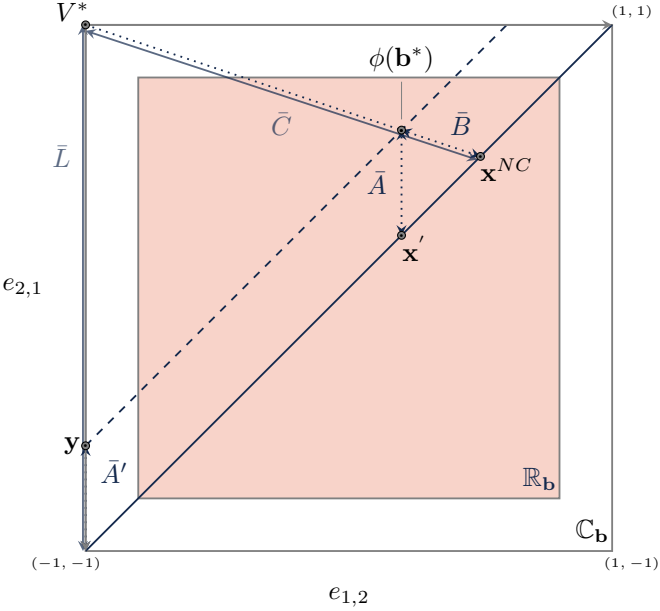

\begin{thm}
If \(\mathcal{R}_{n}\) is a cyclic system of rank \(n\), then \(\text{CNTF} = 2\text{CNT}_{2}\).

\begin{proof}
Let \(\mathcal{R}_{n}\) be a cyclic system of rank \(n\).
Note that if \(\mathcal{R}_{n}\) is noncontextual,
both measures \(\text{CNT}_{2}\) and \(\text{CNTF}\) equal zero and,
trivially, \(\text{CNTF} = 2\text{CNT}_{2}\).
Thus, we will assume that \(\mathcal{R}_{n}\) is contextual.

Without loss of generality, assume that the system \(\mathcal{R}_{n}\)
is consistently connected.
If not, take the consistification \(\mathcal{R}_{n}^{\ddagger}\) of
\(\mathcal{R}_{n}\).
Consider the vector \(\phi(\mathbf{b}^{*})\) and the
corresponding sets \(\mathbb{C}_{\mathbf{b}}\),
\(\mathbb{D}_{\mathbf{b}}\),
\(\mathbb{R}_{\mathbf{b}}\), and
\(\mathbb{E}_{\mathbf{b}}\).
Theorem~15 of \cite{Dzhafarov.2020.Contextuality}
states that
\(\text{CNT}_{2} =
    \frac{1}{4}
    \left\Vert
    \phi(\mathbf{b}^{*}),
    \mathbb{E}_{\mathbf{b}}
    \right\Vert_{1}
\)
is a single coordinate distance.
Since the system is contextual and
consistently connected,
\(
    \left\Vert
    \phi(\mathbf{b}^{*}),
    \mathbb{E}_{\mathbf{b}}
    \right\Vert_{1}
    =
    \left\Vert
    \phi(\mathbf{b}^{*}),
    \mathbb{D}_{\mathbf{b}}
    \right\Vert_{1}
    > 0.
\)
The point \(\phi(\mathbf{b}^{*})\) clearly lies within a
corner formed at an odd vertex of \(\mathbb{E}_{\mathbf{b}}\)
and cut by a facet of \(\mathbb{D}_{\mathbf{b}}\)
(cf. Lemmas 10 and 11 of \cite{Dzhafarov.2020.Contextuality}.
See Figure~\ref{fig:polytope} for an illustration).
Let \(\mathbf{x}^{'}\) be any of the \(n\) points of the facet of
\(\mathbb{D}_{\mathbf{b}}\) that differs from \(\mathbf{b}^{*}\)
on a single coordinate, and
let \(\bar{A}\) denote the line segment between
\(\phi(\mathbf{b}^{*})\) and \(\mathbf{x}^{'}\).
Clearly, the length \(|\bar{A}|\) equals \(4 \text{CNT}_{2}\).

Now, by Theorem~1(iii) of \cite{Abramsky.2017.Contextual},
it is possible to write
\( \phi(\mathbf{b}^{*}) =
    \text{CNTF} \mathbf{x}^{SC} +
    (1 - \text{CNTF}) \mathbf{x}^{NC},
\)
where
\(\mathbf{x}^{NC}\) is a point on the surface of \(\mathbb{D}_{\mathbf{b}}\),
and
\(\mathbf{x}^{SC}\) is a point that maximally violates the
Bell inequality
\begin{equation}
    \label{eq:Bell}
    s_{1}(\mathbf{x}) = \max_{\prod \lambda_{i} = -1}
        \sum_{i = 1}^{n} \lambda_{i} x_{i, i \oplus 1} > n - 2,
\end{equation}
where \(\lambda_{i} = \pm 1\) and the coordinates of \(\mathbf{x}\)
are labeled following the convention to label the coordinates of
\(\phi(\mathbf{b}^{*})\).
Note that the values \(\lambda_{i}^{*}\) for which the maximum is attained
define the coordinates of an odd vertex \(V^{*}\) of
\(\mathbb{C}_{\mathbf{b}}\).
Also, the points \(\mathbf{x}^{SC}\) which maximally violate
(\ref{eq:Bell}) lie in a hyperplane parallel to the facet
of \(\mathbb{D}_{\mathbf{b}}\) facing \(V^{*}\)
and \(V^{*}\) is one of such points\footnote{For
a cyclic system, the odd vertices of \(\mathbb{C}_{\mathbf{b}}\)
are indeed the only points that maximally violate (\ref{eq:Bell})
and are possible points \(\phi(\mathbf{b})\) of a system;
although, they may be unfeasible due to not belonging to
the \(n\)-box \(\mathbb{R}_{\mathbf{b}}\) associated to certain systems.}
(see, e.g., Lemma~11 of \cite{Dzhafarov.2020.Contextuality}).
Thus, one can choose \(\mathbf{x}^{SC} = V^{*}\)
and find \(\mathbf{x}^{NC}\) as the point where
the line that passes through \(V^{*}\)
and \(\phi(\mathbf{b}^{*})\) intersects the corresponding face
of \(\mathbb{D}_{\mathbf{b}}\).
Let \(\bar{B}\) denote the line segment between
\(\phi(\mathbf{b}^{*})\) and \(\mathbf{x}^{NC}\),
and \(\bar{C}\), the line segment between
\(V^{*}\) and
\(\mathbf{x}^{NC}\).
Clearly, \(\text{CNTF} = \nicefrac{|\bar{B}|}{|\bar{C}|}\).

Let \(\mathbf{P}\) be the plane determined by any one edge \(\bar{L}\) of
\(\mathbb{C}_{\mathbf{b}}\) emanating from \(V^{*}\) and segment \(\bar{C}\).
Consider the hyperplanes
\begin{align}
    \label{eq:hypb}
    \sum_{i = 1}^{n} \lambda_{i}^{*} x_{i, i \oplus 1} &=
    \sum_{i = 1}^{n} \lambda_{i}^{*} e_{i, i \oplus 1} \\
    \text{and} & \nonumber \\
    \label{eq:hypface}
    \sum_{i = 1}^{n} \lambda_{i}^{*} x_{i, i \oplus 1} &= n - 2,
\end{align}
Hyperplane~(\ref{eq:hypface}) is the
hyperplane that contains the facet of
\(\mathbb{D}_{\mathbf{b}}\) facing vertex \(V^{*}\),
and hyperplane~(\ref{eq:hypb}) is the hyperplane parallel to (\ref{eq:hypface})
passing through the point \(\phi(\mathbf{b}^{*})\).
The intersections of hyperplanes~(\ref{eq:hypb}) and~(\ref{eq:hypface}) with
\(\mathbf{P}\) determine two parallel lines.
Let \(\mathbf{y}\) be the point at the intersection of
(\ref{eq:hypb}), \(\mathbf{P}\), and \(\bar{L}\).
By Lemma~11 of \cite{Dzhafarov.2020.Contextuality},
the line segment \(\bar{A}'\) between \(\mathbf{y}\) and the end of \(\bar{L}\)
has length \(|\bar{A}|\).
Hence, by Thales' intercept Theorem,
\begin{equation}
    \label{eq:thales}
    \text{CNTF} = \frac{|\bar{B}|}{|\bar{C}|}
                = \frac{|\bar{A}|}{|\bar{L}|}
                = \frac{4 \text{CNT}_{2}}{2}.
\end{equation}
Therefore, \(\text{CNTF} = 2\text{CNT}_{2}\).
\end{proof}
\end{thm}

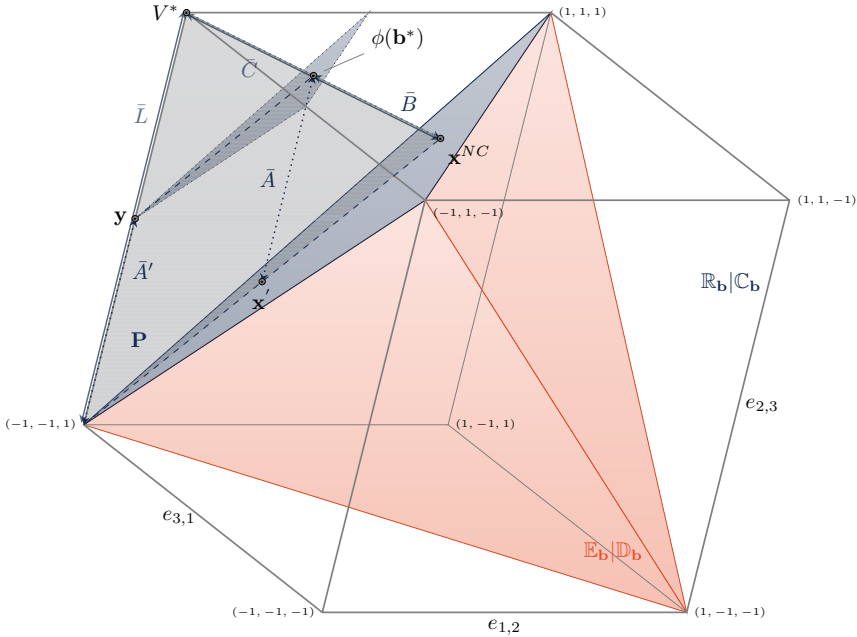
\begin{figure}[tp]

    \caption{Illustration of the proof for a consistently connected system \(\mathcal{R}_{3}\).}

    \begin{center}

   \begin{adjustbox}{max width = .95\textwidth}

\begin{tikzpicture}
\definecolor{naranja}{rgb}{0.90, 0.30, 0.15}
\definecolor{azul}{rgb}{0.08, 0.16, 0.30}

\def\epsilon{.05}

\def \tta{ -14.00000000000000 } 
\def \k{    -3.70000000000000 } 
\def \l{     6.00000000000000 } 
\def \d{     5.00000000000000 } 
\def \h{     7.00000000000000 } 

\coordinate (A) at (0,0);
\coordinate (B) at ({-\h*sin(\tta)},{\h*cos(\tta)});
\coordinate (C) at ({-\h*sin(\tta)-\d*sin(\k*\tta)},
                    {\h*cos(\tta)+\d*cos(\k*\tta)});
\coordinate (D) at ({-\d*sin(\k*\tta)},{\d*cos(\k*\tta)});

\coordinate (Ap) at (\l,0);
\coordinate (Bp) at ({\l-\h*sin(\tta)},{\h*cos(\tta)});
\coordinate (Cp) at ({\l-\h*sin(\tta)-\d*sin(\k*\tta)},
                     {\h*cos(\tta)+\d*cos(\k*\tta)});
\coordinate (Dp) at ({\l-\d*sin(\k*\tta)},{\d*cos(\k*\tta)});

\coordinate (BC) at ({-\h*sin(\tta)-.5*\d*sin(\k*\tta)},
                    {\h*cos(\tta)+.5*\d*cos(\k*\tta)});
\coordinate (CCp) at ({.5*\l-\h*sin(\tta)-\d*sin(\k*\tta)},
                     {\h*cos(\tta)+\d*cos(\k*\tta)});
\coordinate (CD) at ({-.5*\h*sin(\tta)-\d*sin(\k*\tta)},
                    {.5*\h*cos(\tta)+\d*cos(\k*\tta)});

\coordinate (xNC) at ({-(1.15)*\h*sin(\tta)},
                      {(1.15)*\h*cos(\tta)});
\coordinate (phiB) at ({-(2.15)*.5*\h*sin(\tta)-.5*\d*sin(\k*\tta)},
                    {(2.15)*.5*\h*cos(\tta)+.5*\d*cos(\k*\tta)});
\coordinate (xpr) at ({-(1.15)*.5*\h*sin(\tta) - .5*\d*sin(\k*\tta)},
                      {(1.15)*.5*\h*cos(\tta) + .5*\d*cos(\k*\tta)});

\coordinate (Cs) at ({-(\h - .5*\epsilon)*sin(\tta)-\d*sin(\k*\tta)},
                    {(\h - .5*\epsilon)*cos(\tta)+\d*cos(\k*\tta) });
\coordinate (xNCs) at ({-(1.15)*\h*sin(\tta) - .45 *\epsilon},
                      {(1.15)*\h*cos(\tta) - .45 *\epsilon});

\coordinate (Css) at ({-\h*sin(\tta)-(1.01)*\d*sin(\k*\tta)},
                    {\h*cos(\tta)+(1.01)*\d*cos(\k*\tta)});
\coordinate (Ds) at ({-(1.01)*\d*sin(\k*\tta)},{(1.01)*\d*cos(\k*\tta)});

\filldraw[draw=naranja,bottom color=naranja!35, top color=naranja!15]
  (B) -- (D)  -- (Ap) -- (B);
\filldraw[draw=naranja,bottom color=naranja!35, top color=naranja!15]
  (B) -- (Cp) -- (Ap) -- (B);
\filldraw[draw=azul,bottom color=azul!40, top color=azul!25]
  (B) -- (Cp) -- (D) -- (B);

\filldraw[draw=azul,bottom color=azul!40, top color=azul!25, dotted]
  (BC) -- (CCp) -- (CD) -- (BC);

\filldraw[draw=azul,bottom color=azul!30, top color=azul!15, dotted, opacity = .15]
  (D) -- (C) -- (xNC) -- (D);

\draw [gray,-,thin] (Dp) -- (Cp)
                    (Dp) --  (D)
                    (Ap) -- (Dp);

\draw [gray,-,thick] (B)  --  (A)
               (Ap) -- (Bp)
               (B)  --  (C)
               (D)  --  (C)
               (A)  --  (D)
               (Ap) --  (A)
               (Cp) --  (C)
               (Bp) --  (B)
               (Bp) -- (Cp);

\draw [azul,-,dashed] (D)  --  (xNC) ;
\draw [azul,-,dashed] (CD)  --  (phiB) ;

        \draw [line width = .7pt, azul, opacity = .7, stealth-stealth]
            (Cs) -- (xNCs)
            node [below = 3pt, near start] {\(\bar{C}\)};

         \draw [line width = .7pt, azul, dotted]
             (C) -- (phiB);
         \draw [line width = .7pt, azul, dotted, stealth-stealth]
             (phiB) -- (xNC)
            node [above = 2pt, near end] {\(\bar{B}\)};

        \draw [line width = .7pt, azul, opacity = .7, stealth-stealth]
            (Ds) -- (Css)
            node [left = 1pt, near end] {\(\bar{L}\)};

        \draw [line width = .7pt, azul, dotted, stealth-stealth]
             (phiB) -- (xpr)
            node [left = 2pt, midway] {\(\bar{A}\)};
        \draw [line width = .7pt, azul, dotted, stealth-stealth]
             (D) -- (CD)
            node [right = 1pt, near end] {\(\bar{A}'\)};

\draw [gray, thin] (C) -- (xNC) ;
\node [fill = gray, draw, circle, inner sep = 0pt] at (C) {.};
\node [fill = gray, draw, circle, inner sep = 0pt] at (CD) {.};
\node [fill = gray, draw, circle, inner sep = 0pt] at (xNC) {.};
\node [fill = gray, draw, circle, inner sep = 0pt] at (xpr) {.};
\node [fill = gray, draw, circle, inner sep = 0pt] at (phiB) {.};
\node [pin={[pin distance = .7cm]20:\(\phi(\mathbf{b}^{*})\)}] at (phiB) {.};

\draw (Ap) node [right]           {\tiny{\(( 1, -1, -1)\)}}
      (Bp) node [right]           {\tiny{\(( 1,  1, -1)\)}}
      (Cp) node [right]           {\tiny{\(( 1,  1,  1)\)}}
      (C)  node [left,black]      {\(V^{*}\)}
      (D)  node [left]            {\tiny{\((-1, -1,  1)\)}}
      (A)  node [left]            {\tiny{\((-1, -1, -1)\)}}
      (B)  node [below right]     {\tiny{\((-1,  1, -1)\)}}
      (Dp) node [right]           {\tiny{\(( 1, -1,  1)\)}};

\draw (xNC) node [below right] {\(\mathbf{x}^{NC}\)} ;
\draw (xpr) node [below]       {\(\mathbf{x}^{'}\)} ;
\draw (CD)  node [left]        {\(\mathbf{y}\)} ;

         \node [below] at (.5*\l, 0)                                  {\(e_{1,2}\)};
         \node [right] at ({\l-.5*\h*sin(\tta)},{.5*\h*cos(\tta)})    {\(e_{2,3}\)};
         \node [left] at ({-.5*\d*sin(\k*\tta)},{.5*\d*cos(\k*\tta)}) {\(e_{3,1}\)};

         \node [azul, left] at ({\l-.8*\h*sin(\tta)},{.8*\h*cos(\tta)}) {\(\mathbb{R}_{\mathbf{b}} | \mathbb{C}_{\mathbf{b}}\)};

         \node [naranja, above left=+17pt] at (\l, 0.1) {\(\mathbb{E}_{\mathbf{b}} | \mathbb{D}_{\mathbf{b}}\)};

         \node [azul, above right=+10pt] at ({-\d*sin(\k*\tta) + .3},{\d*cos(\k*\tta) + .8}) {\(\mathbf{P}\)};

\end{tikzpicture}

   \end{adjustbox}

   \end{center}
    \label{fig:polytope3d}
    \footnotesize{\raggedright{%
    Note: Example of the noncontextuality polytope and the contextuality
    measures for a consistently connected cyclic system of rank 3 with marginal expectations
    \(e_{i}^{j} =  0\).
    The \(3\)-de\-mi\-cube \(\mathbb{D}_{\mathbf{b}}\) on the even vertices of
    \(\mathbb{C}_{\mathbf{b}}\) is the tetrahedron shaded in orange.
    Due to equal marginal expectations,
    the box \(\mathbb{R}_{\mathbf{b}}\) coincides with the ambient cube \(\mathbb{C}_{\mathbf{b}}\),
    and  the noncontextuality polytope \(\mathbb{E}_{\mathbf{b}}\) coincides
    with \(\mathbb{D}_{\mathbf{b}}\).
    The sections of the (hyper-)planes~(\ref{eq:hypb}) and~(\ref{eq:hypface}) intersecting
    the ambient cube \(\mathbb{C}_{\mathbf{b}}\) are shaded in blue.
    The section of plane \(\mathbf{P}\) between segments \(\bar{C}\) and \(\bar{L}\), and
    the (hyper-)plane (\ref{eq:hypface}) is shaded in gray.
    All other elements as in Figure~\ref{fig:polytope} and described in text.
    }}
\end{figure}

Figure~\ref{fig:polytope} illustrates the construction presented in the proof
for an example consistently connected system of rank 2.
Figure~\ref{fig:polytope3d} provides an example
for a system of rank 3.

\section{Discussion}

The proof presented in this paper improves the understanding of the
relationship between the different approaches to studying contextuality.
\cite{Dzhafarov.2020.Contextuality} showed that
in cyclic systems \(\text{CNT}_{2}\) is equal to another measure, \(\text{CNT}_{1}\),
and in \cite{Kujala.2019.Measures}, it was conjectured that these
measures are proportional to a measure computed based on negative
probabilities, \(\text{CNT}_{3}\).
If this conjecture is found to be true, the current result shows that all these different approaches
to the quantification of contextuality capture essentially the same
feature of cyclic systems.

It should be noted, however, that outside the class of cyclic systems,
these measures are not generally the same.
\cite{Dzhafarov.2020.Contextuality,Dzhafarov.2020.Erratum:}
present some examples that show that \(\text{CNT}_{2}\) and \(\text{CNT}_{1}\)
are not in general functions of each other.
It is also easy to see that \(\text{CNT}_{2}\) and \(\text{CNTF}\) do not need to be proportional
for noncyclic systems.
Take for example any number \(m\) of separate
systems of random variables and consider the system obtained from
their disjoint union.
That is, take systems \(\mathcal{R}_{k}\), with
corresponding sets of contexts and contents \(C_{k}\) and \(Q_{k}\),
\(k = 1, \ldots, m\), and construct the system
\[
\mathcal{R} = \left\{ R_{q}^{c} : c \in \sqcup_{k = 1}^{m} C_{k}, q \in \sqcup_{k = 1}^{m} Q_{k}, q \prec c \right\}.
\]
The following matrix shows an example of such a system, one composed
of two disjoint rank 2 cyclic subsystems.
\begin{equation}
\begin{array}{|c|c|c|c||c|}
\hline
R_{1}^{1} & R_{2}^{1} &           &           & c_{1} \\
\hline
R_{1}^{2} & R_{2}^{2} &           &           & c_{2} \\
\hline
&           & R_{3}^{3} & R_{4}^{3} & c_{3} \\
\hline
&           & R_{3}^{4} & R_{4}^{4} & c_{4} \\
\hline\hline
q_{1}     & q_{2}     & q_{3}     & q_{4}     & \mathcal{R} \\
\hline
\end{array}
\end{equation}
As shown in Cervantes and Dzhafarov \cite{Cervantes.2020.Contextuality},
for these type of systems,
\(\text{CNT}_{2}(\mathcal{R}) = \sum_{k = 1}^{m} \text{CNT}_{2}(\mathcal{R}_{k})\),
whereas
\(\text{CNTF}(\mathcal{R}) = \max_{k = 1,\ldots,m} \text{CNTF}(\mathcal{R}_{k})\).

Lastly, in addition to the measures I have mentioned above, Cervantes
and Dzhafarov \cite{Cervantes.2020.Contextuality} have recently proposed
a hierarchical measure of contextuality and noncontextuality based
on \(\text{CNT}_{2}\). In the class of cyclic systems, this hierarchical
measure reduces to the contextuality measure \(\text{CNT}_{2}\) and
the noncontextuality measure \(\text{NCNT}_{2}\).

\section{Disclosures and acknowledgements}

Declarations of interest: none.

The author carried out
most of
this work as an Illinois
Distinguished Postdoctoral Researcher at the University of Illinois at Urbana-Champaign.

The author is grateful to Sandra Camargo, Ehtibar Dzhafarov and an anonymous reviewer
for feedback on earlier drafts, and to
Ehtibar Dzhafarov and Giulio Camillo for fruitful conversations.
The views and conclusions contained in this document are those of the
author and should not be interpreted as representing the official policies, either expressed
or implied, of the University of Illinois.

\bibliographystyle{unsrt}
\bibliography{CNT2proof.bib}






\appendix

\section{Additional tables and computations for system \(\mathcal{R}_{SQ}\)}
\setcounter{table}{0}

Table~\ref{fig:matrixM} presents the incidence matrix \(\mathbf{M}\) for the contextuality
analysis of system \(\mathcal{R}_{\text{SQ}}\).
Note that the same matrix may be used for the analysis of any cyclic system of rank \(4\).
In Table~\ref{fig:solutionsSQ}, example solutions \(\mathbf{x}^{*}\) and \(\mathbf{z}^{*}\)
to the linear programming tasks
to compute \(\text{CNT}_{2}\) (see Expression~\ref{lp:CNT2}) and
\(\text{CNTF}\) (Expression~\ref{lp:CNTF}), respectively, are presented together with 
the submatrix of \(\mathbf{M}_{(.)}\) whose rows are labeled by the
probabilities of bunch
\(R^{2}_{SQ}\).
The solution \(\mathbf{x}^{*}\) shown in Table~\ref{fig:solutionsSQ}
is such that \(d^{*}_{2}\) is the only \(d^{*}_{i}, i = 1, \ldots, 4,\)
greater than zero.

\begin{sidewaystable}
\caption{Incidence matrix \(\mathbf{M}\) for the contextuality analysis of system
\(\mathcal{R}_{\text{SQ}}\).}

\label{fig:matrixM}

{\tiny{}
\[
\begin{array}{c|c}
\left\{ S_{1}^{1}=1\right\}  & 111111111111111111111111\textbf{1}1111111111111111111111111111111111111111111111111111111111111111111111111111111111111111111111111111111 \\
\left\{ S_{2}^{1}=1\right\}  & 111111111111111111111111\textbf{1}1111111111111111111111111111111111111110000000000000000000000000000000000000000000000000000000000000000 \\
\left\{ S_{2}^{2}=1\right\}  & 111111111111111111111111\textbf{1}1111111000000000000000000000000000000001111111111111111111111111111111100000000000000000000000000000000 \\
\left\{ S_{3}^{2}=1\right\}  & 111111111111111100000000\textbf{0}0000000111111111111111100000000000000001111111111111111000000000000000011111111111111110000000000000000 \\
\left\{ S_{3}^{3}=1\right\}  & 111111110000000011111111\textbf{0}0000000111111110000000011111111000000001111111100000000111111110000000011111111000000001111111100000000 \\
\left\{ S_{4}^{3}=1\right\}  & 111100001111000011110000\textbf{1}1110000111100001111000011110000111100001111000011110000111100001111000011110000111100001111000011110000 \\
\left\{ S_{4}^{4}=1\right\}  & 110011001100110011001100\textbf{1}1001100110011001100110011001100110011001100110011001100110011001100110011001100110011001100110011001100 \\
\left\{ S_{1}^{4}=1\right\}  & 101010101010101010101010\textbf{1}0101010101010101010101010101010101010101010101010101010101010101010101010101010101010101010101010101010 \\
\hline
\left\{ S_{1}^{1}=1,S_{2}^{1}=1\right\}  & 111111111111111111111111\textbf{1}1111111111111111111111111111111111111110000000000000000000000000000000000000000000000000000000000000000 \\
\left\{ S_{2}^{2}=1,S_{3}^{2}=1\right\}  & 111111111111111100000000\textbf{0}0000000000000000000000000000000000000001111111111111111000000000000000000000000000000000000000000000000 \\
\left\{ S_{3}^{3}=1,S_{4}^{3}=1\right\}  & 111100000000000011110000\textbf{0}0000000111100000000000011110000000000001111000000000000111100000000000011110000000000001111000000000000 \\
\left\{ S_{4}^{4}=1,S_{1}^{4}=1\right\}  & 100010001000100010001000\textbf{1}0001000100010001000100010001000100010001000100010001000100010001000100010001000100010001000100010001000 \\
\hline
\left\{ S_{1}^{1}=1,S_{1}^{4}=1\right\}  & 101010101010101010101010\textbf{1}0101010101010101010101010101010101010101010101010101010101010101010101010101010101010101010101010101010 \\
\left\{ S_{2}^{2}=1,S_{2}^{1}=1\right\}  & 111111111111111111111111\textbf{1}1111111000000000000000000000000000000000000000000000000000000000000000000000000000000000000000000000000 \\
\left\{ S_{3}^{3}=1,S_{3}^{2}=1\right\}  & 111111110000000000000000\textbf{0}0000000111111110000000000000000000000001111111100000000000000000000000011111111000000000000000000000000 \\
\left\{ S_{4}^{4}=1,S_{4}^{3}=1\right\}  & 110000001100000011000000\textbf{1}1000000110000001100000011000000110000001100000011000000110000001100000011000000110000001100000011000000 \\
\end{array}
\]
}

{\tiny{}
\[
\begin{array}{c|c}
\left\{ S_{1}^{1}=1\right\}  & 0000000000000000000000000000000000000000000000000000000000000000000000000000000000000000000000000000000\textbf{0}000000000000000000000000 \\
\left\{ S_{2}^{1}=1\right\}  & 1111111111111111111111111111111111111111111111111111111111111111000000000000000000000000000000000000000\textbf{0}000000000000000000000000 \\
\left\{ S_{2}^{2}=1\right\}  & 1111111111111111111111111111111100000000000000000000000000000000111111111111111111111111111111110000000\textbf{0}000000000000000000000000 \\
\left\{ S_{3}^{2}=1\right\}  & 1111111111111111000000000000000011111111111111110000000000000000111111111111111100000000000000001111111\textbf{1}111111110000000000000000 \\
\left\{ S_{3}^{3}=1\right\}  & 1111111100000000111111110000000011111111000000001111111100000000111111110000000011111111000000001111111\textbf{1}000000001111111100000000 \\
\left\{ S_{4}^{3}=1\right\}  & 1111000011110000111100001111000011110000111100001111000011110000111100001111000011110000111100001111000\textbf{0}111100001111000011110000 \\
\left\{ S_{4}^{4}=1\right\}  & 1100110011001100110011001100110011001100110011001100110011001100110011001100110011001100110011001100110\textbf{0}110011001100110011001100 \\
\left\{ S_{1}^{4}=1\right\}  & 1010101010101010101010101010101010101010101010101010101010101010101010101010101010101010101010101010101\textbf{0}101010101010101010101010 \\
\hline
\left\{ S_{1}^{1}=1,S_{2}^{1}=1\right\}  & 0000000000000000000000000000000000000000000000000000000000000000000000000000000000000000000000000000000\textbf{0}000000000000000000000000 \\
\left\{ S_{2}^{2}=1,S_{3}^{2}=1\right\}  & 1111111111111111000000000000000000000000000000000000000000000000111111111111111100000000000000000000000\textbf{0}000000000000000000000000 \\
\left\{ S_{3}^{3}=1,S_{4}^{3}=1\right\}  & 1111000000000000111100000000000011110000000000001111000000000000111100000000000011110000000000001111000\textbf{0}000000001111000000000000 \\
\left\{ S_{4}^{4}=1,S_{1}^{4}=1\right\}  & 1000100010001000100010001000100010001000100010001000100010001000100010001000100010001000100010001000100\textbf{0}100010001000100010001000 \\
\hline
\left\{ S_{1}^{1}=1,S_{1}^{4}=1\right\}  & 0000000000000000000000000000000000000000000000000000000000000000000000000000000000000000000000000000000\textbf{0}000000000000000000000000 \\
\left\{ S_{2}^{2}=1,S_{2}^{1}=1\right\}  & 1111111111111111111111111111111100000000000000000000000000000000000000000000000000000000000000000000000\textbf{0}000000000000000000000000 \\
\left\{ S_{3}^{3}=1,S_{3}^{2}=1\right\}  & 1111111100000000000000000000000011111111000000000000000000000000111111110000000000000000000000001111111\textbf{1}000000000000000000000000 \\
\left\{ S_{4}^{4}=1,S_{4}^{3}=1\right\}  & 1100000011000000110000001100000011000000110000001100000011000000110000001100000011000000110000001100000\textbf{0}110000001100000011000000 \\
\end{array}
\]
}
\raggedright{Note. The matrix is split due to space limitations.
The upper matrix contains the first 128 columns of \(\mathbf{M}\) and
the lower matrix contains the last 128 columns.
The vertical solid
line separates the labels of each row and the matrix block. The horizontal
solid lines split each matrix into the corresponding left and right
blocks of \(\mathbf{M}_{\mathbf{l}}\) (upper block),
\(\mathbf{M}_{\mathbf{b}}\) (middle
block), and \(\mathbf{M}_{\mathbf{c}}\) (lower block).}
\end{sidewaystable}

\begin{sidewaystable}
\caption{Example solutions to the linear programming tasks for computing \(\text{CNT}_{2}\)
and \(\text{CNTF}\) for system \(\mathcal{R}_{\text{SQ}}\) describing
the ``Snow Queen'' experiment.}

\label{fig:solutionsSQ}
\begin{center}
\begin{adjustbox}{max width = 1.0\textwidth}
\begin{tabular}{c|ccccccccccccccccccccccccc}
\(\mathbf{x}^{*\intercal}\) & 0.009 & (5) & 0.018 & (17) & \textbf{0.08} & (95) & 0.029 & (14) & 0.021 & (63) & 0.033 & (31) & \textbf{0.139} & (15) & 0.109 & 0.032 & 0.035 & (1) & 0.36 & (3) & 0.135 \\
\(\mathbf{z}^{*\intercal}\) & 0.009 & (5) & 0.018 & (17) & \textbf{0.011} & (95) & 0.029 & (14) & 0.021 & (63) & 0.033 & (31) & \textbf{0.07} & (15) & 0.109 & 0.032 & 0.035 & (1) & 0.36 & (3) & 0.135 & & \(\mathbf{m}_{\mathbf{prs}}\mathbf{x}^{*}\) & \(\mathbf{m}_{\mathbf{prs}}\mathbf{z}^{*}\) & \(\mathbf{p}^{*}_{(.)}\) \\
\cline{1-22} \cline{24-26}
\(\left\{ S_{2}^{2} = 1, S_{3}^{2} = 1\right\}\) & 1 & (5) & 1 & (17) & \textbf{0} & (59) & 0 & (14) & 1 & (63) & 1 & (31) & \textbf{0} & (15) & 0 & 0 & 0 & (1) & 0 & (3) & 0  & & 0.081 & 0.081 & .150 \\
\(\left\{ S_{2}^{2} = 0, S_{3}^{2} = 1\right\}\) & 0 & (5) & 0 & (17) & \textbf{0} & (95) & 0 & (14) & 0 & (63) & 0 & (31) & \textbf{1} & (15) & 0 & 0 & 0 & (1) & 0 & (3) & 0  & & 0.139 & 0.070 & .070 \\
\(\left\{ S_{2}^{2} = 1, S_{3}^{2} = 0\right\}\) & 0 & (5) & 0 & (17) & \textbf{1} & (95) & 0 & (14) & 0 & (63) & 0 & (31) & \textbf{0} & (15) & 0 & 0 & 0 & (1) & 0 & (3) & 0  & & 0.080 & 0.011 & .011 \\
\(\left\{ S_{2}^{2} = 0, S_{3}^{2} = 0\right\}\) & 0 & (5) & 0 & (17) & \textbf{0} & (95) & 1 & (14) & 0 & (63) & 0 & (31) & \textbf{0} & (15) & 1 & 1 & 1 & (1) & 1 & (3) & 1 & & 0.700 & 0.700 & .769 \\
\cline{1-22} \cline{24-26}
\hline
\end{tabular}
\end{adjustbox}
\end{center}
\raggedright{
\scriptsize
Note. The left panel shows vectors \(\mathbf{x}^{*\intercal}\),
and \(\mathbf{z}^{*\intercal}\).
Vector \(\mathbf{x}^{*}\) solves the
linear programming task for computing \(\text{CNT}_{2}\) (Expression~\ref{lp:CNT2}) so that
\(d_{2} = \text{CNT}_{2}(\mathcal{R}_{\text{SQ}}) = .069\).
Vector \(\mathbf{z}^{*}\) solves the
task to compute \(\text{CNTF}\) (Expression~\ref{lp:CNTF}).
To help readability of the nonzero components, all zero components
of the solutions are omitted and replaced
by the number of omitted elements
inside parentheses.
The lower section shows the
vectors of matrix \(\mathbf{M}_{(.)}\) corresponding to the probabilities of bunch
\(R_{\text{SQ}}^{2}\).
The elements of these vectors corresponding
to the zero components
of the solutions have also been replaced by
the number
of omitted components. The right panel shows the results
of the product of the vectors of matrix \(\mathbf{M}_{(.)}\) and the
solutions to each
linear programming task, as well as the actual probabilities
from bunch \(R_{\text{SQ}}^{2}\).
}
\end{sidewaystable}

\end{document}